\documentclass[runningheads]{llncs}
\usepackage{graphicx}
\usepackage{url}
\usepackage{relsize}
\usepackage{amssymb,amsmath}
\usepackage{microtype}
\usepackage{subcaption}
\usepackage{color}
\usepackage{algorithmicx}
\usepackage[noend]{algpseudocode}
\usepackage[T1]{fontenc}
\usepackage{unixode}
\usepackage{tikz}
\usepackage{hyperref}

\usepackage[subtle]{savetrees}
\usepackage{microtype}

\usetikzlibrary{shapes,arrows,positioning}
\tikzstyle{class}=[draw=black]
\tikzstyle{instance}=[>=stealth]

\usepackage{color}
\definecolor{keywordcolor}{rgb}{0.5, 0.0, 0.0}   
\definecolor{commentcolor}{rgb}{0.4, 0.4, 0.4}   
\definecolor{symbolcolor}{rgb}{0.0, 0.1, 0.6}    
\definecolor{sortcolor}{rgb}{0.0, 0.4, 0.0}      

\usepackage{listings}

\newcommand\ie{\emph{i.e.}\ }
\newcommand\eg{\emph{e.g.}\ }
\newcommand{\icode}[1]{\texttt{\small #1}}

\begin{document}
\title{Tabled Typeclass Resolution}
%
%
\author{Daniel Selsam\inst{1} \and Sebastian Ullrich\inst{2} \and Leonardo de Moura\inst{1}}
\authorrunning{D. Selsam et al.}
%
\institute{Microsoft Research, USA \and Karlsruhe Institute of Technology, Germany}
\maketitle              
\begin{abstract}
Typeclasses provide an elegant and effective way of managing ad-hoc polymorphism
in both programming languages and interactive proof assistants.
However, the increasingly sophisticated uses of typeclasses within proof assistants,
especially within Lean's burgeoning mathematics library, \textsf{mathlib},
have elevated once-theoretical limitations of existing typeclass resolution procedures
into major impediments to ongoing progress.
The two most devastating limitations of existing procedures are exponential running times in the presence of diamonds and divergence in the presence of cycles.
We present a new procedure, \emph{tabled typeclass resolution}, that solves both problems by \emph{tabling},
which is a generalization of memoizing originally introduced to address similar limitations of early logic programming systems.
We have implemented our procedure for the upcoming version (v4) of Lean,
and have confirmed empirically that our implementation is exponentially faster than existing systems in the presence of diamonds.
Although tabling is notoriously difficult to implement,
our procedure is notably lightweight and could easily be implemented in other systems.
We hope our new procedure facilitates even more sophisticated uses of typeclasses in both software development and interactive theorem proving.
\end{abstract}

\section{Introduction}\label{sec:intro}

Typeclasses were introduced in \cite{wadler1989make} as a principled way of enabling \emph{ad-hoc} polymorphism
in functional programming languages,
with the motivating example of overloading equality and arithmetic operators.
They were first implemented in the Haskell programming language \cite{hall1996type},
and since then they have been extended in various ways \cite{chen1992parametric,laufer1996type,jones1997type,jones2000type,kahl2001named,duggan2002type},
have found many diverse uses within Haskell \cite{jones1992computing,gill2000debugging,hinze2001derivable,mcbride2002faking,kiselyov2004strongly,lammel2005scrap,lammel2006software,li2006encoding,pucella2008haskell,keller2010regular,claessen2011quickcheck},
and have emerged as an organizing principle for the language and its libraries \cite{jones2003haskell}.

Typeclasses have also spread to interactive proof assistants such as Isabelle/HOL \cite{nipkow2002isabelle}, Coq \cite{coqv891,sozeau2008first}, and Lean \cite{de2015lean},
and have had a major impact on the organization of libraries of formal mathematics \cite{paulson2004organizing,spitters2010developing,spitters2011type,krebbers2011type,holzl2013type,2019arXiv191009336M}.
Coq also expanded the scope of typeclasses dramatically in \cite{sozeau2008first};
whereas Haskell restricts their use so that resolution can be performed without search,
Coq and Lean both allow much more liberal uses
but at the expense of needing to perform search during typeclass resolution.
Although this tradeoff may or may not prove desirable in more traditional programming languages,
the more flexible typeclass frameworks of Coq and Lean have unlocked remarkable new uses
for typeclasses and are arguably among the key enabling technologies of modern math libraries
such as~\cite{2019arXiv191009336M}.

However, the increasingly sophisticated uses of typeclasses within these contexts have exposed two critical problems with the existing typeclass resolution procedures:
the \emph{diamond problem} and the \emph{cycle problem}.
The diamond problem is that hierarchies of mathematical abstractions contain towers of \emph{diamonds}---\ie multiple paths between
two nodes in the typeclass search tree---and the running time of the existing typeclass resolution procedures on failing subqueries is
exponential in the height of these towers.
The cycle problem is that many desired uses of typeclasses involve cycles in the instance graph, \eg coercing in both directions between two types,
yet such cycles may cause the existing resolution procedures to loop.
Despite the myriad extensions to typeclass mechanisms proposed over the years,
to the best of our knowledge all existing typeclass resolution procedures for interactive proof assistants
are still based on (na\"{i}ve) tree search and are thus susceptible to both exponential blowup in the presence of diamonds and non-termination in the presence of cycles.
A recent paper describing \textsf{mathlib} \cite{2019arXiv191009336M}, the mathematical library of Lean \cite{de2015lean}, highlighted the limitations of the
existing typeclass resolution procedure as one of the main impediments to scaling the library.

\begin{sloppypar}
Similar issues plagued early versions of logic programming systems such as Prolog and Datalog.
Eventually, more sophisticated search procedures were introduced into logic programming systems to address these issues \cite{tamaki1986old,vielle1989recursive,de1994lpda,chen1996tabled,sagonas1998abstract,demoen1998cat,demoen1999chat,guo2001simple,shen2001linear,swift2012xsb,zhou2008linear}.
Although there have been many variants proposed in the literature, most fall under the umbrella of \emph{tabled resolution}.
The main idea of tabled resolution is that during the search, a table is maintained that maps each subgoal to the set of solutions that have already been found for that subgoal,
and solutions are reused from the table whenever possible instead of being recomputed.
Note that na\"{i}ve caching does not suffice, since each subgoal may admit many distinct solutions with dependencies among them.
\end{sloppypar}

In this work, we take inspiration from this line of work and propose a new typeclass resolution procedure that we call \emph{tabled typeclass resolution}
that solves both problems.
Specifically, it eliminates the exponential blowup associated with the diamond problem and guarantees termination under the \emph{bounded term-size assumption} \cite{van1991well}.
  We have implemented our procedure for the upcoming version (v4) of Lean\footnote{\url{http://github.com/leanprover/lean4/blob/IJCAR20/src/Init/Lean/Meta/SynthInstance.lean}},
  and we confirm empirically that our implementation is exponentially faster than existing systems in the presence of diamonds.
Although tabling is notoriously difficult to implement,
our procedure is notably lightweight and could easily be implemented in other systems.
We hope our new procedure facilitates even more sophisticated uses of typeclasses in both software development and interactive theorem proving.

\section{Preliminaries}\label{sec:prelim}

\paragraph{Ad-hoc polymorphism.}
There are two distinct kinds of polymorphism one may want in a programming language: parametric polymorphism and ad-hoc polymorphism \cite{strachey2000fundamental}.
In parametric polymorphism, a function may be defined over a range of types as long as it behaves the same on every type in the range.
For example, one may wish to write a single function \icode{length} that returns the length of an arbitrary list,
no matter what type of element the list contains.

In ad-hoc polymorphism, a function may be defined over a range of types but behave differently on different types in the range.
For example, one may wish to use the same operator \icode{+} to represent addition on many different types, such as natural numbers, rational numbers,
lists and sets, even though the actual implementations of addition are arbitrarily different for each such type. One may also want to write
additional functions like \lstinline{double} in terms of \icode{+} that are agnostic as to how addition is implemented for the type of its argument.

\paragraph{Typeclasses.}
Typeclasses were introduced in \cite{wadler1989make} as a principled way
of enabling ad-hoc polymorphism in functional programming languages.
We first observe that it would be easy to implement
an ad-hoc polymorphic function (such as addition)
if the function simply took the type-specific implementation
of addition as an argument and then called that implementation on the remaining arguments.
For example, suppose we declare a structure in Lean to hold implementations of addition:
\begin{lstlisting}
structure Add (α : Type) := (add : α → α → α)
\end{lstlisting}
Note that this statement is the Lean analogue of the Haskell statement \icode{data Add a = Add \{ add : a -> a -> a \}}. In the above Lean code, the field \lstinline{add} has type
\begin{lstlisting}
add : {α : Type} → Add α → α → α → α
\end{lstlisting}
where the curly braces around the type \lstinline{α} mean that it is an \emph{implicit} argument.
We could implement \lstinline{double} by
\begin{lstlisting}
def double {α : Type} (addα : Add α) (x : α) : α := add addα x x
\end{lstlisting}
and we could double a natural number \lstinline{n} by \lstinline|double { add := natAdd } n|.%
\footnote{Here we assume \lstinline{natAdd} has already been defined.}
Of course, it would be highly cumbersome for users to manually pass the implementations around in this way.
Indeed, it would defeat most of the potential benefits of ad-hoc polymorphism.

The main idea behind typeclasses is to make arguments such as \lstinline{Add α} implicit,
and to use a database of user-defined \emph{instances}
to synthesize the desired instances automatically through a process known as
\emph{typeclass resolution}.
In Lean, by changing \lstinline{structure} to \lstinline{class} in the example above, the type
of \lstinline{add} becomes
\begin{lstlisting}
add : {α : Type} → [Add α] → α → α → α
\end{lstlisting} where the square brackets indicate that the argument
of type \lstinline{Add α} is \emph{instance-implicit}, \ie that it should be synthesized
using typeclass resolution. This version of \lstinline{add} is the Lean analogue of the Haskell term \icode{add :: Add a => a -> a -> a}.
Similarly, we can register an instance by
\begin{lstlisting}
instance natAddInst : Add Nat := { add := natAdd }
\end{lstlisting}
Then for \lstinline{n, m : Nat}, the term \lstinline|add n m| triggers typeclass resolution
with the goal of \lstinline{Add Nat}, and typeclass resolution will synthesize the instance
\lstinline{natAddInst}.
In general, instances may depend on other instances in complicated ways.
For example, we can declare an (anonymous) instance stating that if \lstinline{α} has addition,
then \lstinline{Vec α n} has addition:
\begin{lstlisting}
instance {α} {n : Nat} [Add α] : Add (Vec α n) := { add := addVecAdd }
\end{lstlisting}
The set of instances in a given development can be seen as forming a logic program \cite{robinson1965machine}. For example, the two instances above induce the following two Horn clauses:
\lstinline{Add Nat} and \lstinline{∀ α n, Add α → Add (Vec α n)}.
Given a type \lstinline{T}, terms of type \lstinline{T} constructed using the instances in a development
are in one-to-one correspondence with (resolution) proofs of the corresponding theorem in the induced logic program.
This phenomenon is an instance of the Curry-Howard Isomorphism \cite{curry1934functionality,howard1980formulae}.

\paragraph{Typeclass resolution.}

In Haskell, typeclass resolution does not involve any backtracking search.
Instead, any instance that can resolve with a subgoal is applied eagerly, and
it is an error if there are ever two different instances that can resolve with the same subgoal.
Many systems that employ typeclasses impose similar restrictions, including Isabelle/HOL \cite{nipkow2002isabelle}.
However, many valuable uses of typeclasses (especially in formal mathematics) require deciding among
different ways of synthesizing terms depending on the context.
To support such uses, many systems including Coq, Lean, Agda \cite{bove2009brief}, and Scala \cite{odersky2004overview}
allow overlapping instances and perform backtracking search during typeclass resolution.
In this paper, we only consider this expanded notion of typeclasses for which typeclass resolution requires backtracking search.

Existing typeclass resolution procedures can all be seen as implementing variants of
selective linear definite clause (SLD) resolution \cite{kowalski1974predicate},
which also formed the basis of early Prolog systems.
We defer discussion of the salient differences among existing systems to later sections.
SLD resolution performs a depth-first search of a tree in which every node has an ordered list of remaining subgoals,
and every edge corresponds to resolving the conclusion of a rule against a node's first subgoal.
It maintains a stack of the nodes yet to be expanded, and the main loop is as follows:
\begin{enumerate}
\item Peek at the top node on the stack, with remaining goals \( \overrightarrow{G} \).
\item If \( \overrightarrow{G} \) is empty, the query has been resolved.
\item If all instances have been tried at this node, pop the node and continue.
\item Otherwise, let \( I \) be the next instance not yet tried for this node, and resolve it with the first subgoal \( \mathrm{head}(\overrightarrow{G}) \) to produce new goals \( \overrightarrow{H} \),
  and push a node with subgoals \( \overrightarrow{H} + \mathrm{tail}(\overrightarrow{G}) \) onto the stack.
\end{enumerate}

\paragraph{Typeclasses in interactive proof assistants.}

For our present purposes, the main feature that distinguishes interactive proof assistants from traditional functional programming languages is the ability
to manipulate theorems and proofs in addition to programs.
In particular, in systems such as Lean and Coq, one may define a class that stores not just implementations of functions (\eg \icode{+})
but also proofs about implementations (\eg that \icode{+} is commutative).
There are many valuable uses of typeclasses in interactive theorem proving, such as inferring that types are finite, that predicates are decidable,
and that terms can be coerced into terms of other types.

In this work we focus on the most critical use of typeclasses in formal mathematics: organizing the complex web of relationships between abstract mathematical objects.
For example, informally, a group is a monoid for which the binary operator satisfies additional properties;
a ring has two binary operators, one of which induces a group while the other induces a monoid;
and a field is a ring with both an additional operator and additional properties.
In practice, the web of algebraic relationships is vastly more sophisticated than these informal examples might suggest.
For example, it is critical to distinguish \eg monoids from commutative monoids, groups from abelian groups, rings from semirings,
and fields from division rings. It is also critical to be able to reason about abstract objects together with various orderings,
\eg partially-ordered commutative monoids and linearly-ordered fields.
Decidability must be tracked as well, \eg to distinguish rings with undecidable linear orders from rings with decidable ones.
And of course, the complex web of relationships between all these abstract objects must be maintained as well,
so that \eg a theorem proved about monoids can be used to prove a theorem about groups.

One of the main challenges in building libraries of formal mathematics is organizing these relationships in such a way
that appropriately abstract theorems can be stated, proved, and used conveniently in all appropriate contexts.
Typeclasses have proven to be an elegant and effective way of addressing all of these challenges
and form the basis for prominent libraries of formal mathematics in both Lean and Coq.
However, existing typeclass mechanisms suffer two critical limitations in such regimes.
First, the web of relationships are littered with \emph{diamonds}, \ie multiple ways of showing that one
kind of object is an instance of another, and such diamonds cause exponential blowup in the standard
typeclass resolution procedure. Second, they are also littered with \emph{cycles}, which cause SLD to loop
and so must be carefully preempted. We now describe these problems in more detail.

\section{The Diamond Problem}\label{sec:diamonds}

\begin{figure}
\includegraphics[width=\textwidth]{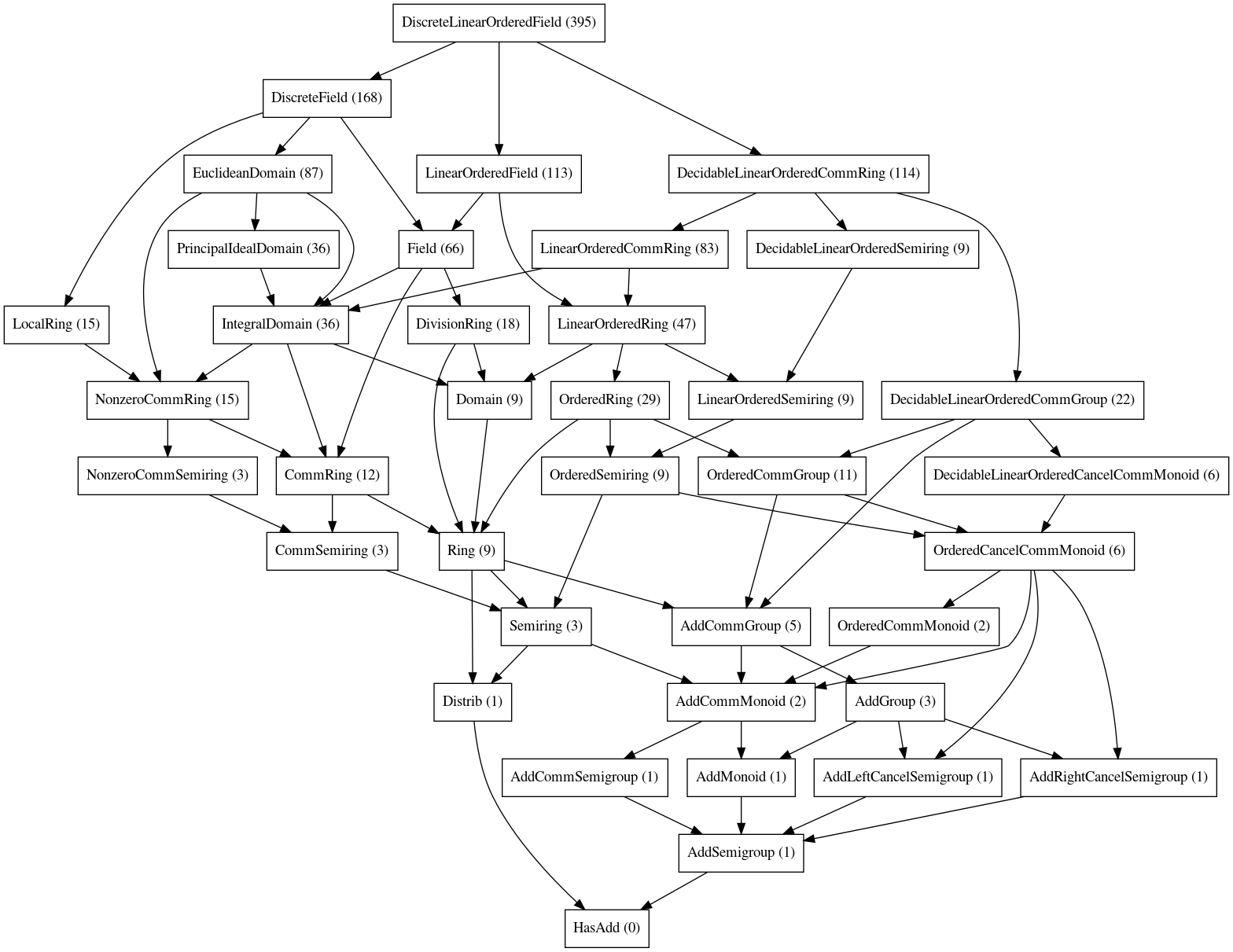}
\caption{A small subgraph of \textsf{mathlib}'s class inheritance graph.
  The rapidly-growing numbers in parentheses indicates the number of distinct paths from a node to the sink, \icode{HasAdd}.}
\label{fig:mathlib-pruned-coercions}
\end{figure}

Diamonds are ubiquitous in formal mathematics.
As mathematician Thomas Hales writes: ``for mathematicians, diamonds are extremely natural and they occur in great abundance under many names (pullbacks, fibered products, Cartesian squares, etc.).'' \cite{halesreview}.
Figure~\ref{fig:mathlib-pruned-coercions} shows a small subgraph of \textsf{mathlib}'s class inheritance graph.
The class inheritance graph consists of the instances automatically generated from subclass declarations,
and forms only a small subset of the instances in \textsf{mathlib}.
Each node in the figure represents a class, and a directed edge from one node to another signifies that the former either inherits directly from the latter,
or that it includes as a parameter an instance of the latter. The number in the parentheses of a node's label indicates how many
distinct paths there are in this subgraph from that node to the sink \lstinline{HasAdd}, which is used to resolve the \lstinline{+} notation.
Crucially, the number of paths to the sink grows exponentially in the height of this graph due to the presence of diamonds.
For example, since there are two paths from \lstinline{AddCommMonoid} and three from \lstinline{AddGroup},
and since \lstinline{AddCommGroup} inherits from both of them, there are five paths from \lstinline{AddCommGroup}.
At the top of the graph, we see that the number of paths nearly triples in a single step, producing almost four hundred paths from \lstinline{DiscreteLinearOrder}.
We stress that this is only (a small subset of) the class inheritance graph, and that most
of the thousands of typeclass instances in \textsf{mathlib} are declared by users
and may introduce arbitrarily complex and context-dependent relationships among classes.
In particular, the sizes and shapes of diamond towers can depend in subtle ways on the both
the set of local instances in scope and the goal type to be synthesized.

The \emph{diamond problem} refers to the exponential blowup that occurs whenever typeclass resolution traverses all paths within a tower of diamonds
of nontrivial depth. As highlighted in \cite{2019arXiv191009336M}, this problem is of severe practical concern in \textsf{mathlib}.
Seemingly innocuous queries may take upwards of ten seconds to succeed,
and may involve traversing the same tower of diamonds upwards of twenty-five thousand times.\footnote{Data included in supplementary material.}
The exponential running times can happen for two related reasons in SLD resolution:
an entire tower may fail, or a tower may succeed
but downstream goals may fail thus causing all solutions to the tower to be enumerated in sequence.

To the best of our knowledge, all existing typeclass resolution procedures take exponential time on the first case.
We show empirically in \S\ref{sec:experiments} that this is true of Lean3, Coq, Agda, and Scala.
We note that while Lean3's procedure is effectively vanilla SLD and so takes exponential time in both cases,
Coq extends SLD in such a way that it avoids exponential work in (a common version of) the second case.
Specifically, Coq detects when a subgoal neither appears in any downstream goals nor contains unification variables in its type, in which
case Coq's procedure commits to the first solution found for that subgoal rather than na\"{i}vely enumerating alternative solutions that are considered unlikely to affect failing
downstream goals.\footnote{Note that the logics of Coq and Lean are so expressive that even for a given type with no unification variables,
  different choices of instances may indeed cause different downstream goals to succeed and fail;
however, it is common practice in both systems to ignore this possibility and to assume that instances are ``morally canonical''.}

\section{The Cycle Problem}\label{sec:cycles}

A second problem of the typeclass resolution procedures based on SLD resolution is that it may not (and without care, does not) terminate,
even when the number of distinct subgoals encountered is finite. This shortcoming imposes severe limitations on the use of typeclasses.
For example, the original typeclass paper \cite{wadler1989make} suggested a \lstinline{Coe} typeclass that represents coercions from type \lstinline{α} to \lstinline{β}:
\begin{lstlisting}
class Coe (α β : Type) : Type := (coe : α → β)
\end{lstlisting}
The idea is that if a term \lstinline{(x : α)} of type \lstinline{α} is used in a context expecting a term of type \lstinline{β},
then typeclass resolution would try to synthesize a term of type \lstinline{Coe α β} and replace \lstinline{(x : α)} with \lstinline{(coe x : β)}.
This is indeed how coercions are managed in Lean. However, the natural instance encoding transitivity would introduce a trivial cycle with a new useless subgoal
added at every step:
[\lstinline{Coe α β}] $\to$ [\lstinline{Coe α ?x₁}, \lstinline{Coe ?x₁ β}] $\to$ [\lstinline{Coe α ?x₂}, \lstinline{Coe ?x₂ ?x₁}, \lstinline{Coe ?x₁ β}], and so on.
There is a known workaround for this limitation: represent the transitive closure of \lstinline{Coe} in a separate class \lstinline{CoeT},
and define the transitivity instance to take one instance of type \lstinline{Coe ?x₁ ?x₂} (to be tried first) and one instance of type
\lstinline{CoeT ?x₂ ?x₃}.
However, even with this workaround, coercing between two types in both directions would still cause SLD resolution to loop.
Thus even though \eg finite sets and finite multisets may usefully coerce into each other, when using SLD resolution,
one direction must be chosen arbitrarily for the \lstinline{Coe} instance and the other must be sacrificed.

Cycles can be very convenient in many areas of formal mathematics, and Lean3's failure to handle them has
been a frequent source of frustration for \textsf{mathlib} users. One desirable instance allows restricting the scalars in a module:
\begin{lstlisting}
class Module (A M : Type) [Ring A] [AddCommGroup M] : Type
class Algebra (R A : Type) [CommRing R] [Ring A] : Type
instance {k A M : Type} {c : CommRing k} {r : Ring A}
         {g : AddCommGroup M} [Algebra k A] [Module A M] : Module k M
\end{lstlisting}
This instance would immediately loop if the \lstinline{Module A M} subgoal were tried first,
but can also induce a loop even if \lstinline{Algebra k A} were tried first,
\eg due to an existing instance that every commutative ring is an algebra over itself.

The \emph{cycle problem} refers to the infinite loops that occur when typeclass resolution blindly tries to solve
a particular goal as a subgoal of itself. There are two cases worth distinguishing: when there is no solution to the query,
and when there is a solution that may be missed due to the loop.
Lean3's procedure is effectively vanilla SLD, and it loops in both cases.
Coq's procedure loops in the first case, but it can be made to succeed in the second case by toggling the iterative-deepening flag.
We note that GHC supports certain types of cycles as well,
\eg instances with recursive constraints that can be resolved coinductively (without search)~\cite{lammel2005scrap}
and recursive superclass declarations for which expansion may only terminate for subtle reasons~\cite{recursivesuperclasses}.
However, most of the cycles we are interested in---including the two examples just discussed---do not fall into these fragments
and seem to require search to resolve.

\section{Tabled Typeclass Resolution}\label{sec:ttr}

We now describe our new typeclass resolution procedure, \emph{tabled typeclass resolution},
that avoids the exponential blowup resulting from towers of diamonds and that
guarantees termination under the \emph{bounded term-size assumption} \cite{van1991well}.
Our procedure is based on the tabled resolution procedure introduced for Prolog in \cite{sagonas1998abstract}.
We have implemented our procedure for the upcoming version (v4) of Lean.
Our actual implementation supports some advanced features that may not be feasible in all relevant languages,
but these features are orthogonal to the new procedure itself and so we focus our presentation on the universally-applicable core.

\subsection{High-level description}
\label{sec:resolution:overview}
Recall from \S\ref{sec:prelim} that SLD resolution performs a depth-first search of a tree in which every node has an ordered list of remaining subgoals,
and every edge corresponds to resolving the conclusion of a rule against a node's first subgoal.
Most of the problems with SLD resolution arise from the fact that SLD will try to solve the same subgoals over and over again in different
parts of the search tree.
Whereas SLD resolution maintains a single search tree for the entire resolution problem,
tabled typeclass resolution maintains a search \emph{forest}, with a distinct search tree for each distinct subgoal
(up to $\alpha$-equivalence) encountered during resolution.
Whenever a subgoal is encountered,
rather than searching for solutions to it from scratch as in SLD,
tabled typeclass resolution looks for the search tree corresponding to that subgoal (in the so-called ``table'').
If the search tree does not already exist, the current branch of the search forest is \emph{suspended},
and control jumps to the new search tree.
On the other hand, if the search tree does already exist,
and if there are already solutions for it, those solutions are used and the search continues.

There are several other cases that need to be considered as well.
For example, the search tree may already exist, but there may not be any solutions to it yet,
in which case control does not jump to the search tree, but the fact that the current
branch of the search forest depends on it is still recorded.
Whenever a new solution is found to any search tree, all other branches of the search forest that depend on it
are \emph{resumed} with the new solution.
Thus the search may be highly \emph{nonlinear}. Indeed, although the algorithm we present is relatively simple,
it can nonetheless induce sophisticated and counterintuitive control flow.

Our tabled typeclass resolution procedure distinguishes between two types of nodes: \emph{generator nodes} and \emph{consumer nodes}.
Generator nodes are in one-to-one correspondence with the search trees, and form the roots of these trees.
A generator node behaves like the root node of the SLD search tree, in the sense that it stores a list of instances to try,
and it is expanded by resolving its subgoal with these instances in sequence.
We use an indexing datastructure to map subgoals to the list of instances to try, and we discuss this further in \S\ref{sec:resolution:indexing}.
All other nodes are \emph{consumer nodes}, and like
the internal nodes in SLD, consumer nodes maintain a list of subgoals that remain to be solved to establish the
subgoal corresponding to its search tree (which we refer to as its \emph{ancestor goal}). However, in contrast to SLD nodes, consumer nodes
are not expanded by resolving their first subgoal against the available instances,
but rather by resolving it against the \emph{solutions} that have been found for the search tree corresponding
to that subgoal.

Our tabled typeclass resolution procedure maintains two distinct stacks,
the \emph{generator stack} for generator nodes, and the \emph{resume stack} for (solution, consumer node) pairs
that have yet to be tried. It also maintains a \emph{table}, which maps each distinct subgoal to
a \emph{table entry} that includes the set of solutions already discovered for it
along with every consumer node that is known to depend on it.

\begin{figure}
  \small
\begin{algorithmic}[1]
  \Procedure{TabledTypeclassResolution}{query}
  \State \Call{newSubgoal}{query}\label{ttcr:gnode:root}
  \While{true}
  \If{resume stack is not empty}\label{ttcr:resumestacknotempty}
  \State pop \icode{(cnode, solution)} from resume stack \label{ttcr:poppair}
  \If{first subgoal of \icode{cnode} does not resolve with \icode{solution}} \icode{continue} \EndIf \label{ttcr:tryresolve}
  \If{\icode{cnode} has no remaining subgoals} \label{ttcr:noremaining}
  \State extract new solution \icode{s} to \icode{cnode}'s ancestor goal \icode{g} \label{ttcr:extract}
  \If{\icode{g} is original query} \Return \icode{s} \EndIf \label{ttcr:solved}
  \If{\icode{s} already in table} continue \EndIf
  \State add \icode{s} to cnode's table entry \label{ttcr:newsolution}
  \For{every cnode \icode{c} dependent on \icode{g}}
  \State push \icode{(c, s)} onto resume stack \label{ttcr:resume}
  \EndFor
  \Else \State \Call{newConsumerNode}{remaining subgoals} \label{ttcr:newconsumernode1}
  \EndIf
  \ElsIf{generator stack is not empty} \label{ttcr:genstacknotempty}
  \State peek at \icode{gnode} on top of generator stack \label{ttcr:peekgstack}
  \If{no remaining instances for \icode{gnode} to try} pop generator stack \EndIf \label{ttcr:popgstack}
  \If{next instance resolves with \icode{gnode}'s goal} \label{ttcr:trynext} \State \Call{newConsumerNode}{new subgoals} \label{ttcr:newconsumernode2}
  \EndIf
  \Else \State \icode{fail} \label{ttcr:fail}
  \EndIf
  \EndWhile
  \EndProcedure
  \Procedure{newSubgoal}{subgoal}
  \State insert new table entry for subgoal into table \label{ttcr:inserttable}
  \State find instances that might resolve with subgoal \label{ttcr:findinsts}
  \State push new generator node for subgoal onto generator stack \label{ttcr:pushgnode}
  \EndProcedure
  \Procedure{newConsumerNode}{subgoals}
  \If{first subgoal \icode{g} of subgoals is not in table} \Call{newSubgoal}{\icode{g}} \label{ttcr:newsubgoal} \EndIf
  \State for each solution to \icode{g}, push new cnode onto resume stack with it \label{ttcr:resumequeued}
  \State add new cnode to \icode{g}'s dependents \label{ttcr:newdep}
  \EndProcedure
\end{algorithmic}
\caption{High-level pseudocode for tabled typeclass resolution.}
  \label{proc:outerloop}
\end{figure}

Figure~\ref{proc:outerloop} provides high-level pseudocode for tabled typeclass resolution.
Before entering the main loop, it registers the query as a new subgoal (\ref{ttcr:gnode:root}),
which involves creating a new table entry for it (\ref{ttcr:inserttable}), finding the list of instances to try for it (\ref{ttcr:findinsts}),
and then pushing a new generator node for it onto the generator stack (\ref{ttcr:pushgnode}).
Then with the resume stack still empty, tabled typeclass resolution enters its main loop.

In the main loop, it first checks to see if the resume stack is nonempty (\ref{ttcr:resumestacknotempty}).
If it is, it pops a pair from it (\ref{ttcr:poppair}), and tries resolving the first subgoal of the popped consumer node
with the popped solution (\ref{ttcr:tryresolve}).
If the unification succeeds and if the consumer node has no remaining subgoals (\ref{ttcr:noremaining}), then a new solution to the consumer node's ancestor goal has been discovered (\ref{ttcr:extract}).
If the ancestor goal happens to be the original query, it returns the solution (\ref{ttcr:solved}).
Otherwise, if the solution is new, it must be added to the corresponding table entry (\ref{ttcr:newsolution}),
and all other consumer nodes that depend on the ancestor goal must be pushed onto the resume stack along with the newly discovered solution (\ref{ttcr:resume}).
If the unification succeeds but the consumer node still has subgoals remaining,
then its creates a new consumer node with them (\ref{ttcr:newconsumernode1})
by calling \Call{NewConsumerNode}{} on the list of subgoals.
This subroutine first checks to see if the first subgoal has been visited yet, and if it has not, it registers a new subgoal for it (\ref{ttcr:newsubgoal}).
Then, for each existing solution to the first subgoal, it pushes the consumer node along with that solution onto the resume stack (\ref{ttcr:resumequeued}), and finally registers the fact that the new consumer node depends on its first subgoal (\ref{ttcr:newdep}).

On the other hand, if the resume stack is empty but the generator stack is not (\ref{ttcr:genstacknotempty}),
then it instead peeks at the generator node on the top of the generator stack (\ref{ttcr:peekgstack}).
If there are no remaining instances to be tried, it pops the generator node and continues (\ref{ttcr:popgstack}).
If there are still instances to be tried then it tries the next instance (\ref{ttcr:trynext}), and if it succeeds, creates a new consumer node for the remaining subgoals (\ref{ttcr:newconsumernode2}).
If both stacks are empty, the procedure fails (\ref{ttcr:fail}).

\subsection{Example}

\begin{figure}
  \begin{center}
  \includegraphics[width=\textwidth]{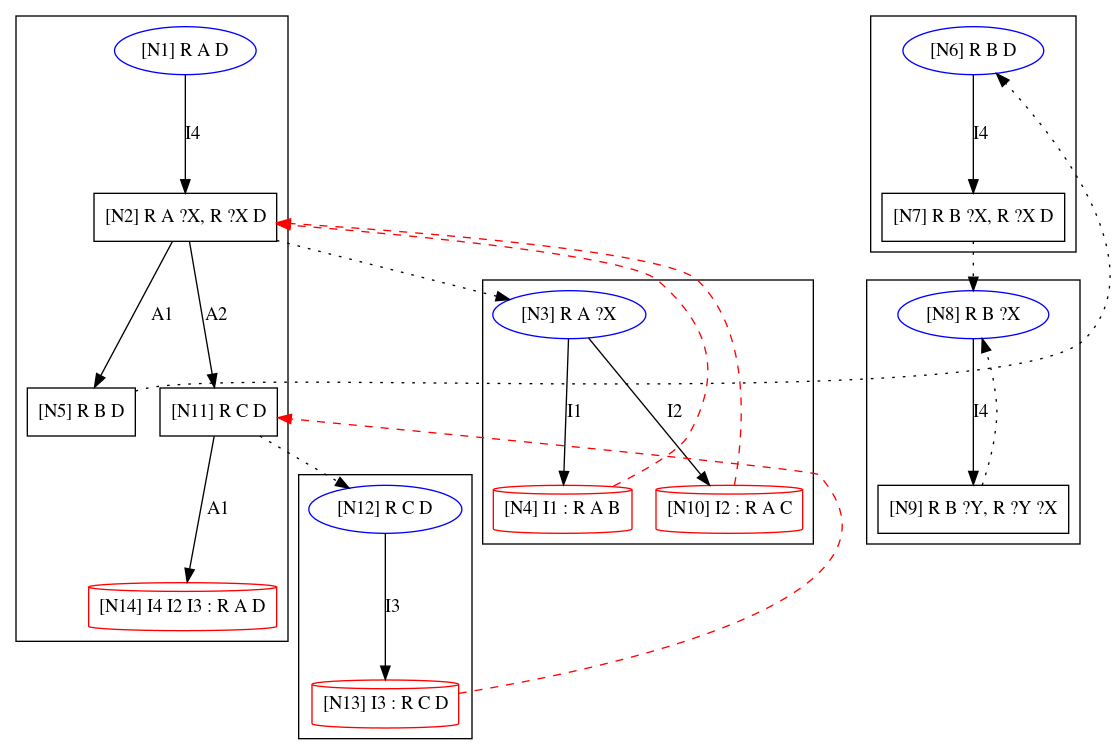}
  \end{center}
  \caption{
    Illustration of tabled typeclass resolution running on the example problem.
    Blue (ellipse) nodes indicate generator nodes, black (rectangle) nodes indicate consumer nodes, and red (cylinder) nodes indicate solutions.
    The nodes are numbered in the sequence that they are created.
    Each distinct search tree corresponding to a distinct subgoal is contained in its own rectangle.
    The edges within trees are solid and represent the resolution of the first subgoal of a node with either an instance
    or a solution from the table.
    There are two types of edges between trees: black (dotted) edges that represent a dependence of a consumer node on a subgoal,
    and red (dotted) edges indicating a solution being used to resume a consumer node.
  }
  \label{fig:example:graph}
\end{figure}

Before discussing implementation details,
we first provide more intuition for our procedure by walking through the following small example:
\begin{lstlisting}
instance I1 : R A B
instance I2 : R A C
instance I3 : R C D
instance I4 {X Y Z : Type} : R X Y → R Y Z → R X Z
#synth R A D -- call typeclass resolution on goal `R A D`
\end{lstlisting}
In this example, a transitive relation \lstinline{R} satisfies the three ground facts \lstinline{R A B}, \lstinline{R A C} and \lstinline{R C D},
and the goal is to synthesize a term of type \lstinline{R A D}.
Figure~\ref{fig:example:graph} shows a visualization of tabled typeclass resolution running on this example.

\begin{table}
  \begin{center}
   \begin{tabular}{|l|l|l|}
   \hline
   Subgoal & Solutions & Dependents \\
   \hline
   \icode{R A D}  & \icode{I4 I2 I3 : R A D}               &     \\
   \icode{R A ?X} & \icode{I1 : R A B}, \icode{I2 : R A C} & N2  \\
   \icode{R B D}  &                                        & N5  \\
   \icode{R B ?X} &                                        & N7, N9 \\
   \icode{R C D} & \icode{I3 : R C D}                      & N11 \\
   \hline
   \end{tabular}
   \end{center}
   \caption{The state of the table when tabled resolution finishes on the example problem.}
   \label{table:example:table}
\end{table}

We first create a generator node (N1) for the original goal, \lstinline{R A D}.
We also create a new table entry for \lstinline{R A D},
since it is the first time this subgoal has been encountered.
The final table entries at the end of the procedure are shown in Table~\ref{table:example:table}.
Next, we try to resolve \lstinline{R A D} with the instances in sequence.
The instance \lstinline{I4} is the only one that succeeds, and it produces the two subgoals \lstinline{R A ?X} and \lstinline{R ?X D}.
We create a new consumer node (N2) with these two subgoals, and with N1 as parent.
Since we have not encountered any $\alpha$-variant of N2's first subgoal \lstinline{R A ?X} yet,
we create a new generator node (N3) and a new table entry for it.
We add N2 to N3's dependency list to indicate that N2 will eventually consume
the solutions found for the subgoal of N3.
This dependency is represented in Figure~\ref{fig:example:graph} by the dotted line leaving N2.
Next, we expand the generator node N3 by trying to resolve its subgoal with the instances in sequence.
\lstinline{I1} succeeds and produces the solution \lstinline{I1 : R A B}. We add this solution (N4)
to the table entry for \lstinline{R A ?X}, and then resume N2 with it.
Resolving N2's first subgoal \lstinline{R A ?X} with \lstinline{R A B} yields
a new consumer node (N5) with \lstinline{R B D} as its only subgoal. We have not encountered \lstinline{R B D} yet,
so we create a new generator node (N6) for it as well as a new table entry, and add N5 to its dependency list.

The procedure proceeds as we have just described until creating consumer node N9,
whose first subgoal \lstinline{R B ?Y} is an $\alpha$-variant of the subgoal \lstinline{R B ?X} that already has a table entry.
N9 is added to the dependency list for that subgoal, and since there are no solutions in the table yet for \lstinline{R B ?X}, control backtracks
to the most recent generator node with instances that have not been tried yet (in this case N3),
and continues the search with the next instance.
Eventually, node N14 constitutes a solution to the original query, which the resolution procedure returns.

\subsection{Suspending and resuming branches}
Tabled resolution procedures for Prolog were heavily complicated by the need
to save and restore the environments (\ie the current assignment of unification variables)
whenever suspending and resuming branches of the search forest.
To resume a branch, the entire sequence of variable assignments from the root to the current node
needed to be replayed.
We did not even mention this challenge in the pseudocode of \S\ref{sec:resolution:overview}
because our procedure does nothing special to support saving or resuming environments:
it simply stores the environment for each node using \emph{persistent data structures}
which enable compact storage of overlapping environments as well as
constant-time copies (see \cite{kaplan2018persistent} for an overview).
The usual downside of using persistent datastructures is that querying, inserting, and deleting
are generally slower than in their imperative counterparts.
For workloads with many queries between backtracking steps (as might arise inside an SMT solver \cite{barrett2018satisfiability},
the performance overhead of persistent datastructures may be devastating.
However, typeclass resolution generally has the opposite profile:
frequent, non-linear context jumps with relatively few queries at each step.
Since saving and restoring environments using persistent datastructures are both constant time,
persistent datastructures may even provide better performance characteristics for typeclass resolution than their imperative counterparts.
Moreover, they \emph{dramatically} simplify the implementation.

\subsection{Finding equivalent subgoals}
\label{sec:resolution:details:table}
The table of distinct subgoals forms a key part of our tabled typeclass resolution procedure.
However, a simple map datastructure does not suffice, since the operations need to be performed modulo $\alpha$-equivalence.
The standard approach for implementing the table in tabled resolution is to use a discrimination tree,
as in \cite{sagonas1998abstract}. Although this approach would work for us as well, our implementation
simply $\alpha$-normalizes the subgoal and then uses a regular hash map on the normalized result.
Specifically, before performing any map operation, we traverse the type of the subgoal and replace all unassigned unification variables with
constants with a reserved prefix (say \lstinline{θ}) and ascending integer suffixes. For example,
\lstinline{f ?X (g ?Y ?X)} and \lstinline{f ?Y (g ?Z ?Y)} would both be normalized to \lstinline{f θ₀ (g θ₁ θ₀)},
and so would map to the same subgoal in our table.

There are pros and cons to the two approaches. The discrimination tree approach has an advantage if subgoals
tend to have many variables in them, since looking up an existing subgoal in the table modulo $\alpha$-equivalence can be done without any allocations.
On the other hand, the approach we take has an advantage if subgoals tend to have few variables in them, because
each subterm can store a single bit indicating the presence of a unification variable, and $\alpha$-normalization can short-circuit on all
ground subterms. In contrast, inserting into a discrimination tree will always require a linear traversal over the subgoal.
Notably, our approach also avoids the quadratic blowup usually associated with tabled resolution.
The classic example from Prolog is the ternary predicate \lstinline{Append} that
computes the concatenation of its first two arguments and stores it in its third argument.
Our procedure can resolve such queries in (quasi-)linear time for two reasons.
In the \lstinline{Append} example, the third argument is always a variable,
while the first two are large but variable-free, and so our short-circuiting $\alpha$-normalization takes constant time.

Lastly, we note that while in traditional logic programming, the only relevant form of equivalence is $\alpha$-equivalence,
there are many other forms of equivalence in expressive logics such as intensional type theory (which forms the basis of both Lean and Coq),
including $\beta$, $\iota$, $\delta$, $\eta$, and $\zeta$.
Our approach can be made to operate modulo additional forms of equivalence by simply performing additional reductions during the normalization step.

\subsection{Indexing the instances}
\label{sec:resolution:indexing}
As mentioned in \S\ref{sec:resolution:overview}, each generator node stores a list of instances to try.
There may be thousands of instances, yet in general only a small number of them will resolve successfully with a given subgoal.
Thus, it is valuable to use an indexing datastructure to map subgoals to small supersets of the instances that may resolve with them.
Discrimination trees and related datastructures work well for this problem in first-order logic~\cite{sekar2001term},
but the problem is much harder and less well studied for more expressive logics such as intensional type theory
in which unification is performed modulo certain types of reductions.
Our approach is similar to the one used in Coq: store the instances in a discrimination tree,
and expose user-facing options that affect which terms are treated as rigid (at the expense of returning under-approximations)
and which are treated as wildcards (at the expense of returning over-approximations).

\subsection{Additional considerations}
\label{sec:resolution:details}
\paragraph{Nested typeclass resolution.}
Many typeclass resolution queries in \textsf{mathlib} require solving nested typeclass resolution problems during unification.
For example, consider the following toy Lean snippet:
\begin{lstlisting}
class B (α : Type) : Type
class L (α : Type) : Type
class R (α : Type) : Type
class F (α : Type) [B α] : Type
instance LtB {α : Type} [L α] : B α
instance RtB {α : Type} [R α] : B α
instance RtL {α : Type} [R α] : L α
instance LtF {α : Type} [L α] : F α
#synth (α : Type) [r : R α], F α
\end{lstlisting}
When showing implicit arguments, the goal \lstinline{F α} is \lstinline{@F α (@RtB α r)},
and the conclusion of \lstinline{LtF} is \lstinline{@F ?α (@LtB ?α ?l)}.
Resolving the former with the latter produces the unification subproblem \lstinline{@RtB α r =?= @LtB ?α ?l},
which requires synthesizing a term of type \lstinline{?l : L α} in order to solve.
Triggering typeclass resolution inside the unifier to solve goals of this form can be seen as the
analogue of unification hints for canonical structures \cite{mahboubi2013canonical}.
Lean3 supported this feature as well, and \textsf{mathlib} relies on it extensively.
To the best of our knowledge, Coq does not yet support this feature.

\paragraph{Scheduling strategies.}
The pseudocode we presented in \S\ref{sec:resolution:overview} uses a very simple scheduling strategy,
and we have found this simple approach to work well for us empirically.
However, many other scheduling policies have been proposed in the tabled resolution literature (see \eg \cite{swift2012xsb})
all of which could be applied in our regime as well.

\paragraph{Incremental garbage collection.}
The nonlinear control flow of tabled resolution makes incremental garbage collection trickier than in SLD resolution.
In particular, when a generator node is popped from the generator stack, it does not mean that all solutions to its subgoal have been discovered,
since there may still be consumer nodes from the same search tree that are suspended on other not-yet-exhausted subgoals.
An efficient way of detecting exhausted subgoals is presented in \cite{sagonas1998abstract}, and the same approach would work in our setting as well.
However, we have found our memory usage to be negligible even on sophisticated queries,
and so do not consider incremental garbage collection to be worth the cost in our regime.

\section{Experiments}
\label{sec:experiments}
\begin{figure}
\centering
  \begin{subfigure}[b]{0.47\textwidth}
    \includegraphics[width=\textwidth]{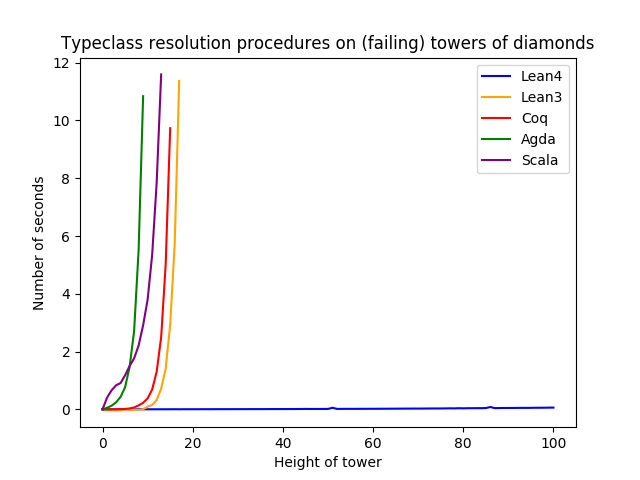}
    \caption{}
    \label{fig:experiments:diamond}
  \end{subfigure}
  \begin{subfigure}[b]{0.47\textwidth}
    \includegraphics[width=\textwidth]{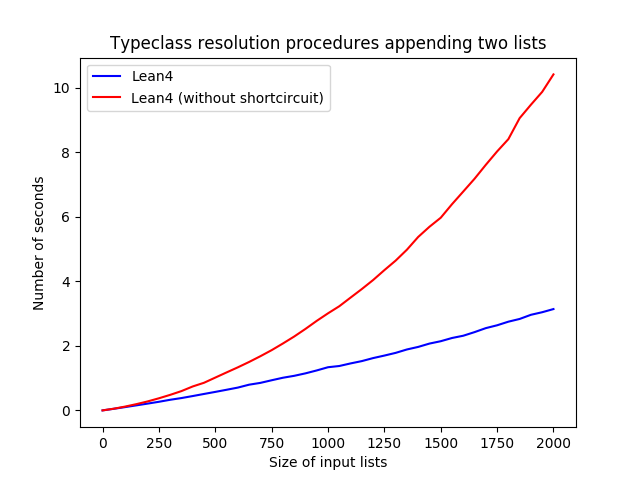}
    \caption{}
    \label{fig:experiments:append}
  \end{subfigure}
  \caption{Comparing the performance of typeclass resolution procedures on failing towers of diamonds (\ref{fig:experiments:diamond})
    and appending two lists (\ref{fig:experiments:append}).}
  \label{fig:experiments}
\end{figure}

Unfortunately, it is not feasible to translate realistic \textsf{mathlib} queries
from Lean3 into other systems---not even Lean4---due to a panoply of critical differences among them.
Thus, we settle for empirically validating the main asymptotic claims made in the paper
on two classes of synthetic typeclass resolution problems.\footnote{Reproducible code included in supplementary material.}
We first evaluate on the quintessential (failing) tower of diamonds:
\begin{lstlisting}
instance BtL (α : Type) (n : ℕ) [B α n] : L α n
instance BtR (α : Type) (n : ℕ) [B α n] : R α n
instance LtT (α : Type) (n : ℕ) [L α n] : T α n
instance RtT (α : Type) (n : ℕ) [R α n] : T α n
instance TtB (α : Type) (n : ℕ) [T α n] : B α (n+1)
#synth T Unit n -- (for some n)
\end{lstlisting}
where \lstinline{B}, \lstinline{L}, \lstinline{R}, \lstinline{T} stand for ``bottom'', ``left'', ``right'', and ``top'' respectively.
Figure~\ref{fig:experiments:diamond} shows the performance of the various systems on this problem as a function of the \lstinline{n}
in the query \lstinline{T Unit n}.
We see that our new tabled resolution procedure (Lean4) is indeed exponentially faster than
existing procedures.
Second, we confirm empirically in Figure~\ref{fig:experiments:append} that our approach to subgoal indexing discussed in~\S\ref{sec:resolution:details:table} indeed
avoids the quadratic blowup associated with tabling on the classic \lstinline{Append} example.

\section{Conclusion}
The increasingly sophisticated uses of typeclasses within proof assistants,
especially within Lean's burgeoning mathematics library, \textsf{mathlib},
have exposed two critical problems with the existing typeclass resolution procedures:
exponential running times in the presence of diamonds and divergence in the presence of cycles.
We have presented a new procedure, \emph{tabled typeclass resolution}, that solves both problems by \emph{tabling},
which is a generalization of memoizing originally introduced to address similar limitations of early logic programming systems.
We have implemented our procedure for the upcoming version (v4) of Lean,
and have confirmed empirically that our implementation is exponentially faster than existing systems in the presence of diamonds.
Although tabling is notoriously difficult to implement,
our procedure is notably lightweight and could easily be implemented in other systems.
We hope our new procedure facilitates even more sophisticated uses of typeclasses in both software development and interactive theorem proving.

\paragraph{Acknowledgments.}
We thank the \textsf{mathlib} community for extensively stress-testing Lean3 and
in so doing, exposing the problems that we have addressed in this paper.
We give special thanks to Kevin Buzzard, Floris van Doorn, Johan Commelin, Patrick Massot, Mario Carneiro,
Rob Lewis, S\'{e}bastien Gou\"{e}zel, Chris Hughes, Yury Kudryashov,
Scott Morrison, Johannes H\"{o}lzl, Oliver Nash, Jason Rute, Reid Barton, and Jeremy Avigad for helping
us to understand the problematic idioms and to isolate representative examples.
We also thank Richard Eisenberg, Simon Peyton Jones, Rob Lewis, Johan Commelin and Marc Huisinga for helpful feedback on early drafts.

\bibliographystyle{splncs04}
\bibliography{typeclass}
\end{document}